\documentstyle[preprint,aps]{revtex}

\tighten
\begin{document}

\title{
       No-core shell-model calculations with 
       starting-energy-independent multi-valued effective
       interactions
}
\medskip

\author{
        P. Navr\'atil\footnote{On the leave of absence from the
   Institute of Nuclear Physics,
                   Academy of Sciences of the Czech Republic,
                   250 68 \v{R}e\v{z} near Prague,
                     Czech Republic.}
     and B. R. Barrett
        }

\medskip

\address{
                   Department of Physics,
                   University of Arizona,
                   Tucson, Arizona 85721
}

\maketitle

\bigskip

\begin{abstract}
Large-space no-core shell model calculations have been performed
for $^3$H, $^4$He, $^5$He, $^6$Li, and $^6$He, using a 
starting-energy-independent two-body effective interaction derived by
application of the
Lee-Suzuki similarity transformation.
This transformation can be performed by direct calculation
or by different iteration procedures,
which are described.
A possible way of reducing the auxiliary potential influence 
on the two-body effective interaction has also been introduced.
The many-body effects have been partially taken into account by 
employing the
recently introduced multi-valued effective interaction approach. 
Dependence of the $^5$He energy levels 
on the harmonic-oscillator frequency as well as on the size of the
model space has been studied. 
The Reid 93 nucleon-nucleon potential
has been used in the study, but results have also been obtained 
using the Nijmegen II potential for comparison.
\end{abstract}

\bigskip
\bigskip
\bigskip

\narrowtext



\section{Introduction}
\label{sec1}

Large-basis no-core shell-model calculations have recently been 
performed \cite{ZBJVC,JZBV,JHBV,ZBVC,ZB94,ZVB,ZBVM,ZBVHS}. 
In these calculations all nucleons are active,
which simplifies the effective interaction as no hole states
are present. In the approach taken, the effective interaction is
determined for a system of two nucleons only and subsequently
used in the many-particle calculations. To take into account a
part of the many-body effects a multi-valued effective interaction 
approach was introduced and applied in the no-core shell-model 
calculations \cite{ZBVHS}
and also tested in a model calculation \cite{NB96}. 

In these shell-model calculations different 
approaches have been taken in deriving the two-nucleon effective
interaction. All the methods employed, however, relied on the
reference G-matrix method, introduced in Ref. \cite{BHM71},
which leads to two-body matrix elements of a starting-energy-dependent
G-matrix.
To get rid of
this unwanted dependence either a suitable parametrization was chosen
\cite{ZBVC,ZB94,ZVB,ZBVHS} or folded diagram 
effects were taken into account
by calculating the derivatives of the G-matrix in an approximate 
way \cite{ZBVC,ZBVM}.

In the present paper we apply the Lee-Suzuki similarity-transformation
approach \cite{LS80} to derive the two-body effective interaction. We try
to avoid unnecessary approximations in performing the calculations. 
The harmonic-oscillator insertions are kept; consequently,
the effective interaction is $A$-dependent. Also the hermitization
of the effective interaction, which is, in general, 
non-hermitian, is done without
approximations by a similarity transformation. 

We consider three
possible ways of deriving the effective interaction. The first
two are the standard iterative procedures, starting with
the $G$-matrix calculation \cite{BHM71}. 
We study the Lee-Suzuki vertex renormalization iteration,
which makes use of the $G$-matrix derivatives. Unlike most of 
the previous applications of this approach, 
we calculate these derivatives exactly employing
the reference $G$-matrix derivatives. The other iterative procedure,
usually refered to as Krenciglowa-Kuo technique, 
is carried out by diagonalizing the
non-hermitian effective hamiltonian in subsequent iterations \cite{KK}. 
This method has so far been applied only to model calculations.
We also discuss its recent generalization \cite{KKSO}. Finally,
we directly construct the effective interaction 
without calculating the $G$-matrix by employing 
the solutions of the two-body
problem. When criteria necessary for convergence are the same 
for both iteration procedures, the resulting effective interactions
obtained in all three methods are identical. 
The last method, however, has two advantages; 
first, its simplicity, and second,
the fact that it utilizes an explicit construction
of the transformation operator.
This transformation operator may then be used for 
the calculation of other effective operators.  

To take partially into account the many-body effects
neglected when using only a two-body effective interaction,
we employ the recently introduced multi-vauled effective
interaction approach \cite{ZBVHS}.

It was observed earlier \cite{ZBVM} 
that the two-body effective interaction
derived by the vertex renormalization method is too attractive and
leads to overbinding of the many-body system. We discuss this problem
and believe that it is largely due to the uncompensated $Q$ space
part of the auxilliary harmonic-oscillator potential. We discuss 
possible treatment of this problem by modification of that part
of the auxilliary potential.

In section \ref{sec2} we discuss the shell model hamiltonian
with the bound center-of-mass as well as the two-particle hamiltonian
and the methods used to derive the starting-energy-independent
effective interaction. Results of the calculations for  
$^3$H, $^4$He, $^5$He, $^6$Li, and $^6$He are presented in
section \ref{sec3}. In particular, we discuss the 
harmonic-oscillator frequency and the model-space-size
dependences of the $^5$He states. 
Conclusions are given in section \ref{sec4}.

\section{Shell model hamiltonian and starting-energy independent
effective interaction}
\label{sec2}

In most shell model studies the one- and two-body hamiltonian
for the A-nucleon system, i.e.,
\begin{equation}\label{ham}
H=\sum_{i=1}^A \frac{\vec{p}_i^2}{2m}+\sum_{i<j}^A V_{ij} \; ,
\end{equation}
where $m$ is the nucleon mass and $V_{ij}$ the nucleon-nucleon interaction,
is modified by adding the center-of-mass harmonic oscillator potential
$\frac{1}{2}Am\Omega^2 \vec{R}^2$, 
$\vec{R}=\frac{1}{A}\sum_{i=1}^{A}\vec{r}_i$.
This potential does not influence intrinsic properties of the 
many-body system. It provides, however, a mean field felt by each nucleon
and allows us to work with a convenient harmonic oscillator basis.
For an alternative manipulation of the center-of-mass terms see, e.g.,
Ref. \cite{ZBJVC}.
The modified hamiltonian, depending on the harmonic oscillator 
frequency $\Omega$, can be written as
\begin{equation}\label{hamomega}
H^\Omega=\sum_{i=1}^A \left[ \frac{\vec{p}_i^2}{2m}
+\frac{1}{2}m\Omega^2 \vec{r}^2_i
\right] + \sum_{i<j}^A \left[ V_{ij}-\frac{m\Omega^2}{2A}
(\vec{r}_i-\vec{r}_j)^2
\right] \; ,
\end{equation}
which is the same as Eq. (4) in Ref. \cite{ZBJVC}. 
Shell-model calculations are carried out in a model space defined
by a projector $P$. In the present work we will always use a complete 
$N\hbar\Omega$ model space. The complementary space to the model space
is defined by the projector $Q=1-P$. Consequently, for the $P$-space
part of the shell-model hamiltonian we get
\begin{equation}\label{phamomega}
H^\Omega_P=\sum_{i=1}^A P\left[ \frac{\vec{p}_i^2}{2m}
+\frac{1}{2}m\Omega^2 \vec{r}^2_i
\right]P + \sum_{i<j}^A P\left[ V_{ij}-\frac{m\Omega^2}{2A}
(\vec{r}_i-\vec{r}_j)^2
\right]_{\rm eff} P \; .
\end{equation}
The effective interaction appearing in Eq.(\ref{phamomega}) is, in general,
an A-body interaction, and, if it is determined without any approximations,
the model-space hamiltonian provides an identical description 
of a subset of states as the full-space hamiltonian (\ref{hamomega}).
The intrinsic properties of the many-body system still do not depend
on $\Omega$. From among the eigenstates of the hamiltonian 
(\ref{phamomega}),
it is necessary to choose only those corresponding to the same 
center-of-mass energy. This can be achieved by projecting 
the center-of-mass eigenstates
with energies greater than $\frac{3}{2}\hbar\Omega$ upwards in the
energy spectrum. The shell-model hamiltonian, which does this, 
takes the form
\begin{eqnarray}\label{phamomegabeta}
H^\Omega_{P\beta}=\sum_{i<j=1}^A &P&\left[ 
\frac{(\vec{p}_i-\vec{p}_j)^2}{2Am}
+\frac{m\Omega^2}{2A} (\vec{r}_i-\vec{r}_j)^2
\right]P + \sum_{i<j}^A P\left[ V_{ij}-\frac{m\Omega^2}{2A}
(\vec{r}_i-\vec{r}_j)^2
\right]_{\rm eff} P  \nonumber \\
&+& \beta P(H^\Omega_{\rm cm}-\frac{3}{2}\hbar\Omega)P\; ,
\end{eqnarray}
where $\beta$ is a sufficiently large positive parameter and 
$H^\Omega_{\rm cm}=\frac{\vec{P}_{\rm cm}^2}{2Am}
+\frac{1}{2}Am\Omega^2 \vec{R}^2$,
$\vec{P}_{\rm cm}=\sum_{i=1}^A \vec{p}_i$. 
In a complete $N\hbar\Omega$ model space
the removal of the spurious center-of-mass motion is exact.
When going from Eq.(\ref{phamomega}) to (\ref{phamomegabeta}),
we added $(\beta-1)PH^\Omega_{\rm cm}P$ and subtracted 
$\beta\frac{3}{2}\hbar\Omega P$, which has no effect
on the intrinsic spectrum of states with the lowest 
center-of-mass configuration.

The effective interaction should be determined from $H^\Omega$ 
(\ref{hamomega}). Calculation of the exact A-body effective
interaction is, however, as difficult as finding the full space solution.
Usually, the effective interaction is approximated by a 
two-body effective interaction determined from a two-nucleon
problem. The relevant two-nucleon hamiltonian 
obtained from (\ref{hamomega}) is then
\begin{equation}\label{hamomega2}
H^\Omega_2\equiv H^\Omega_{02}+V_2^\Omega=
\frac{\vec{p}_1^2+\vec{p}_2^2}{2m}
+\frac{1}{2}m\Omega^2 (\vec{r}^2_1+\vec{r}^2_2)
+ V(\vec{r}_1-\vec{r}_2)-\frac{m\Omega^2}{2A}(\vec{r}_1-\vec{r}_2)^2 \; .
\end{equation}
This can be transformed to two-nucleon relative and center-of-mass
parts by introducing the coordinates 
$\vec{q}=\frac{1}{2}(\vec{p}_1-\vec{p}_2)$,
$\vec{P}_{\rm 2cm}=\vec{p}_1+\vec{p}_2$, 
$\vec{R}_{\rm 2cm}=\frac{1}{2}(\vec{r}_1+\vec{r}_2)$
and $\vec{r}=\vec{r}_1-\vec{r}_2$ yielding
\begin{equation}\label{hamomega2r}
H^\Omega_2=\frac{\vec{P}_{\rm 2cm}^2}{2M}+\frac{1}{2}M\Omega^2 
\vec{R}_{\rm 2cm}^2
+\frac{\vec{q}^2}{2\mu}+\frac{A-2}{2A}\mu\Omega^2 \vec{r}^2
+ V(\vec{r}) \; ,
\end{equation}
with $M=2m$ and $\mu=\frac{1}{2}m$. While the center-of-mass
part has the solution $E_{\cal NL}=(2{\cal N}+{\cal L}
+\frac{3}{2})\hbar\Omega$
and the eigenvectors $|{\cal NL}\rangle$, the relative-coordinate part
can be solved as a differential equation or, alternatively,
can be diagonalized in a sufficiently large harmonic oscillator
basis. The latter possibility is, obviously, not applicable 
for hard-core potentials. 

The starting-energy-dependent effective interaction 
or G-matrix corresponding to a two-nucleon model
space defined by the projector $P_2$ can be written as
\begin{equation}\label{G}
G(\varepsilon) = V_2^\Omega 
+ V_2^\Omega Q_2 \frac{1}{\varepsilon-Q_2H_2^\Omega Q_2} 
Q_2 V_2^\Omega \; ,
\end{equation}
where $Q_2=1-P_2$ and $V_2^\Omega$ is the interaction given by the last 
two terms on the rhs of Eq.(\ref{hamomega2}). The G-matrix (\ref{G}) can
be constructed from the solutions of the Schr\"{o}dinger equation
with the hamiltonian (\ref{hamomega2}), by using the reference
matrix method \cite{BHM71}. The G-matrix can be expressed as
\begin{mathletters}\label{GGR}\begin{eqnarray}
G(\varepsilon) &=& A(\varepsilon)^{-1} G_{\rm R}(\varepsilon) \; ,
\label{GAGR}\\
G_{\rm R}(\varepsilon)&=&(H_{02}^\Omega-\varepsilon)+
(H_{02}^\Omega-\varepsilon)\sum_{k}\frac{|k\rangle\langle k|}
{\varepsilon -E_k}(H_{02}^\Omega-\varepsilon) \; , \label{GR} \\
A(\varepsilon)&=&1+G_{\rm R}(\varepsilon) P_2 \frac{1}
{\varepsilon-H_{02}^\Omega} \; , \label{Aeps} 
\end{eqnarray}\end{mathletters}
where  $H_{02}^\Omega$ is given by the first two terms on the rhs 
of Eq.(\ref{hamomega2})
and $E_k$, $|k\rangle$ are the eigenvalues and eigenvectors of
$H_2^\Omega$ (\ref{hamomega2}), respectively. 

To obtain a starting-energy-independent effective interaction,
one has to take into account the folded diagrams or, equivalently, 
to construct a similarity transformation that guarantees
decoupling between the model space $P$  and the Q-space. 
We employ the Lee-Suzuki \cite{LS80} similarity transformation
method, which gives the effective interaction in the form
\begin{equation}\label{LS}
P_2V_{\rm 2eff}P_2 = P_2V_2^\Omega P_2 
+ P_2V_2^\Omega Q_2\omega P_2 \; ,
\end{equation}
with $\omega$ satisfying the equations $\omega=Q_2\omega P_2$ and
\begin{equation}\label{omega}
\omega=Q_2\frac{1}{\varepsilon-Q_2H_2^\Omega Q_2}Q_2V_2^\Omega P_2 
-Q_2\frac{1}{\varepsilon-Q_2H_2^\Omega Q_2}
Q_2\omega P_2 (H_2^\Omega +H_2^\Omega Q_2\omega-\varepsilon)P_2 \; .
\end{equation}
In this degenerate-model-space formulation two iterative 
solutions of the equation (\ref{omega}) exist and lead to different
expressions for the effective interaction. The first one, 
the Krenciglowa-Kuo (KK) iteration procedure \cite{KK},
gives the for the n-th iteration formula
\begin{equation}\label{KKit}
V_{{\rm 2eff},n} = \sum_l P_2 G(\varepsilon+E_{l,n-1})P_2|l_{n-1}
\rangle\langle\tilde{l}_{n-1}|P_2 \; .
\end{equation}
In (\ref{KKit}) the states $|l_{n-1}\rangle$ are the right 
eigenvectors
of $H_{02}^\Omega+V_{{\rm 2eff},n-1}-\varepsilon$ belonging 
to the eigenvalue $E_{l,n-1}$. The tilda states are the 
biorthogonal eigenvectors.
When this procedure converges, it does so to the states 
which have the largest
overlap with the model-space states. 
The resulting $V_{\rm 2eff}$ (\ref{LS}) is independent
of the starting energy $\varepsilon$. 
This method was recently generalized to a non-degenerate 
model space \cite{KKSO}. The difference is in the starting 
iteration. When using (\ref{KKit}), the starting iteration
is $G(\varepsilon)$, while the non-degenerate model 
space version \cite{KKSO} starts with 
\begin{equation}\label{GDelta}
\sum_\alpha G(\varepsilon_\alpha+\Delta)P_{2\alpha} \; .
\end{equation}
Here the $\varepsilon_\alpha$ are the unperturbed energies, in our case
the two-nucleon harmonic-oscillator energies, 
and $P_{2\alpha}$ is the projector
on the two-nucleon state $|\alpha\rangle$. Unlike in the original paper
\cite{KKSO}, we introduce a shift $\Delta$, as the G-matrix cannot
be evaluated for the energy $\varepsilon_\alpha$, using the reference
matrix method (\ref{GGR}). Note that (\ref{GDelta}) is the interaction
used in the previous no-core shell-model studies 
\cite{ZBVC,ZB94,ZVB,ZBVHS} with $\Delta$ treated as a free parameter.

One can also use the alternative
procedure, usually called the vertex-renormalization approach,
which can be obtained from Eq.(\ref{omega}) \cite{LS80}.
The resulting iteration sequence for the effective interaction
can be written as
\begin{eqnarray}\label{verren}
H_{02}^\Omega+V_{{\rm 2eff},n}-\varepsilon &=&
\left[P_2-P_2 G^{(1)}(\varepsilon) P_2\right.
\nonumber \\
&&-\sum_{m=2}^{n-1} \frac{1}{m!}
P_2 G^{(m)}(\varepsilon) P_2
(H_{02}^\Omega+V_{{\rm 2eff},n-m+1}-\varepsilon)P_2
(H_{02}^\Omega+V_{{\rm 2eff},n-m+2}-\varepsilon)P_2
\nonumber \\ 
&&\left.
\ldots (H_{02}^\Omega+V_{{\rm 2eff},n-1}-\varepsilon)P_2 \right]^{-1} 
P_2 \left(H_{02}^\Omega+G(\varepsilon)-\varepsilon\right)P_2 \; ,
\end{eqnarray}
where $V_{{\rm 2eff},1}=P_2G(\varepsilon)P_2$.
This method, which relies on the G-matrix derivatives $G^{(m)}$, 
was applied in shell-model calculations
for the first time by Poppelier and Brussard \cite{PB91}. 
In that and most other applications different
numerical approximations were used to evaluate the derivatives
\cite{ZBVC,ZBVM,PB91}. In fact, these derivatives can be calculated 
exactly, because the reference matrix $G_{\rm R}(\varepsilon)$ 
(\ref{GR}) as well as the operator $A(\varepsilon)$ (\ref{Aeps})
can be differentiated to any order analytically, as can be seen
when they are expressed in the two-nucleon harmonic-oscillator basis.
For example, the $n$-th derivative, $n>0$, of $G_{\rm R}$ is
\begin{eqnarray}\label{GRder}
G_{{\rm R}\alpha\gamma}^{(n)}(\varepsilon)&=& (-1)^n n!\left[
\sum_{k}\frac{\langle\alpha|k\rangle\langle k|\gamma\rangle}
{(\varepsilon -E_k)^{n-1}}-(2\varepsilon-\varepsilon_\alpha
-\varepsilon_\gamma)
\sum_{k}\frac{\langle\alpha|k\rangle\langle k|\gamma\rangle}
{(\varepsilon -E_k)^n} \right. \nonumber \\
 &+&\left. (\varepsilon-\varepsilon_\alpha)
\sum_{k}\frac{\langle\alpha|k\rangle\langle k|\gamma\rangle}
{(\varepsilon -E_k)^{n+1}}(\varepsilon-\varepsilon_\gamma)
\right] \; ,
\end{eqnarray}
and similarly for $A(\varepsilon)$. Then the $n$-th G-matrix
derivative can be expressed as
\begin{equation}\label{Gder}
G^{(n)}=\sum_{m=0}^n \frac{n!}{(n-m)!m!} (A^{-1})^{(n-m)}
G_{\rm R}^{(m)} \; ,
\end{equation}
where the derivatives of the inverse matrix can be obtained
from
\begin{equation}\label{Ainvder}
(A^{-1})^{(n)}=-\sum_{m=0}^{n-1} \frac{n!}{(n-m)!m!} A^{-1}A^{(n-m)}
(A^{-1})^{(m)} \; .
\end{equation}
An alternative way of calculating G-matrix derivatives analytically
was proposed in Ref. \cite{MWC93}.
When the convergence of (\ref{verren}) is achieved, the effective 
interaction does not depend on the starting energy $\varepsilon$.
The states reproduced in the model space are those lying closest
to $\varepsilon$.

As our calculations start with exact solutions of the hamiltonian
(\ref{hamomega2}), we are, in fact, in  a position to construct
the operator $\omega$ and, hence, the effective interaction directly
from these solutions. Let us denote the two-nucleon 
harmonic-oscillator states, which form the model space, 
as $|\alpha_P\rangle$,
and those which belong to the Q-space, as $|\alpha_Q\rangle$.
Then the Q-space components of the eigenvector $|k\rangle$ of
the hamiltonian (\ref{hamomega2}) can be expressed as a combination
of the P-space components with the help of the operator $\omega$
\begin{equation}\label{eigomega}  
\langle\alpha_Q|k\rangle=\sum_{\alpha_P}
\langle\alpha_Q|\omega|\alpha_P\rangle \langle\alpha_P|k\rangle \; .
\end{equation}
If the dimension of the model space is $d_P$, we may choose a set
${\cal K}$ of $d_P$ eigenevectors, 
for which the relation (\ref{eigomega}) 
will be satisfied. Under the condition that the $d_P\times d_P$ 
matrix $\langle\alpha_P|k\rangle$ for $|k\rangle\in{\cal K}$
is invertible, the operator $\omega$ can be determined from 
(\ref{eigomega}). Note that in the present application the eigenvectors
$|k\rangle$ are direct products of the center-of-mass and the relative
coordinate eigenvectors. The condition of invertibility is not
satisfied for an arbitrary choice of the eigenvector set ${\cal K}$.
Once the operator $\omega$ is determined the effective hamiltonian
can be constructed as follows from Eq.(\ref{LS}) 
\begin{equation}\label{effomega}
\langle \gamma_P|H_{2\rm eff}|\alpha_P\rangle =\sum_{k\in{\cal K}}
\left[
\langle\gamma_P|k\rangle E_k\langle k|\alpha_P\rangle
+\sum_{\alpha_Q}\langle\gamma_P|k\rangle E_k\langle k|\alpha_Q\rangle
\langle\alpha_Q |\omega|\alpha_P\rangle\right] \; .
\end{equation}
It should be noted that 
$P_2|k\rangle=\sum_{\alpha_P}|\alpha_P\rangle\langle\alpha_P|k\rangle$
for $|k\rangle\in{\cal K}$ is a right eigenvector of (\ref{effomega})
with the eigenvalue $E_k$.

In the case when the iteration conditions are the same for both
methods (\ref{KKit}) and (\ref{verren}), and the set ${\cal K}$,
for which Eq.(\ref{eigomega}) is fulfilled, is chosen accordingly, all
three methods lead to the identical effective hamiltonian. This 
hamiltonian, when diagonalized in a model space basis, reproduces
exactly the set ${\cal K}$ of $d_P$ eigenvalues $E_k$. Note that
the effective hamiltonian is, in general, non-hermitian, or
more accurately quasi-hermitian. It can be hermitized by a similarity
transformation. When the direct method 
(\ref{eigomega},\ref{effomega}) is used the similarity transformation
is determined from the metric operator $P_2(1+\omega^\dagger\omega)P_2$. 
The hermitian hamiltonian is then given by \cite{S82SO83}
\begin{equation}\label{hermeffomega}
\bar{H}_{\rm 2eff}
=\left[P_2(1+\omega^\dagger\omega)P_2\right]^{\frac{1}{2}}
H_{\rm 2eff}\left[P_2(1+\omega^\dagger\omega)
P_2\right]^{-\frac{1}{2}} \; .
\end{equation}
When the iteration methods (\ref{KKit}) or (\ref{verren}) are employed,
the $\omega$ operator is not determined. In previous applications
\cite{ZBVC,ZBVM,PB91}, 
the hermitization was usually done by averaging the conjugate matrix
elements. However, also in this case  
a similarity transformation can be constructed, which hermitizes the
effective hamiltonian. As shown in Ref. \cite{SGH}, 
if $S$ is a matrix, which diagonalizes
$H_{\rm 2eff}$, $SH_{\rm 2eff}S^{-1}=D$, then the metric operator
can be expressed as $S^\dagger S$. Consequently, it follows 
that the hermitized $\bar{H}_{\rm 2eff}$ is obtained from 
\begin{equation}\label{hermeff}
\bar{H}_{\rm 2eff}=\left[S^\dagger S\right]^{\frac{1}{2}}
H_{\rm 2eff}\left[S^\dagger S\right]^{-\frac{1}{2}}   \; .
\end{equation}

Finally, the two-body effective interaction used 
in the shell-model calculation
is determined from the two-nucleon effective hamiltonian 
as $V_{\rm 2eff}=\bar{H}_{\rm 2eff}-H_{02}^\Omega$.

\section{Application to light nuclei}
\label{sec3}

In this section we apply the methods for calculating
the two-body effective interaction outlined in section \ref{sec2}
and, with the obtained interactions, we perform no-core shell-model
calculations for nuclei with $A=3-6$. We use a complete $N\hbar\Omega$
model space with, e.g., $N=8$ for the positive-parity states of $^4$He.
This means that 9 major harmonic-oscillator shells may be occupied
in this case. The two-nucleon model space is defined in our calculations
by $N_{\rm max}$, e.g., for an $8\hbar\Omega$ calculation for $^4$He
$N_{\rm max}=8$. 
The restriction of the harmonic-oscillator shell occupation is 
given by 
$N_1\le N_{\rm max}$, $N_2\le N_{\rm max}$, $(N_1+N_2)\le N_{\rm max}$.
The same conditions hold for the relative $2n+l$, 
center-of-mass $2{\cal N}+{\cal L}$, and
$2n+l+2{\cal N}+{\cal L}=N_1+N_2$, quantum numbers.

In the present calculations we use the Reid 93 nucleon-nucleon
potential \cite{SKTS}. For $^4$He we make a comparison with the   
Nijmegen II potential \cite{SKTS} with corrected $^1{\rm P}_1$
wave \cite{Spc}. We work in the isospin formalism and
do not include the Coulomb interaction.
The np channel interactions of Reid 93 and Nijmegen II are used.

First, let us compare the different methods for calculating
the starting-energy-independent effective interaction.
The KK method (\ref{KKit}) works well for $N_{\rm max}\le 4$.
Then the convergence is achieved usually after less than
15 iterations and the model space eigenvalues differ from
the full space eigenvalues, which are of the order of 
$10^1-10^2$ MeV, by not more than $10^{-4}$ MeV. Note that
the reference G-matrix (\ref{GR}) becomes singular for
$\varepsilon=E_k$. Consequently, exact eigenvalues cannot
be reproduced when the reference G-matrix method is applied
for calculating the G-matrix. However, this singularity
causes no problem when the iteration is stopped after
achieving the above mentioned precision. When the starting
iteration (\ref{GDelta}) is used instead of $G(\varepsilon)$,
usually a few iteration steps are saved. As to the starting
energy, $\varepsilon=0$ is a possible choice. In most
calculations we used negative values for $\varepsilon$. 
For $\Delta$
a non-zero value must be chosen, we used typically about -5 MeV.
The resulting effective interaction is not dependent on these
choices, and they also do not affect the number of iterations
significantly. For $N_{\rm max}> 4$ divergence in some channels
can be encountered and, moreover, the calculation becomes
time consuming. 

The application of the vertex-renormalization method (\ref{verren})
requires the G-matrix derivatives. The matrix elements of 
the G-matrix derivatives decrease rapidly but, on the other hand,
they are multiplied by effective-interaction matrices of
increasing powers. Consequently, the overall convergence is
rather slow for larger model spaces. 
Also the rate of convergence differs for different
states. 
When the starting energy is chosen below the ground state
energy, the fastest convergence is achieved 
for the lowest states, 
which have the largest overlap with the model space.
This is the same observation as found 
in the model calculations \cite{NG93}.
The rate of convergence is very sensitive to the choice of the
starting energy $\varepsilon$. The closer to the ground state,
the faster the convergence. On the other hand, serious numerical
problems occur when a higher number of iterations is required.
These problems are related to the above mentioned fact that
we multiply very small numbers by very big numbers when
calculating  higher iterations. To achieve the same accuracy
as with the previous method, often more than 30 iterations are
needed, and to curb the numerical difficulties real*16 precision
is neccessary. These facts make this method rather impractical.
It is also difficult to apply this method for model spaces 
with $N_{\rm max}>4$.  

Let us point out that new iteration methods were suggested recently,
which combine some features of both methods studied here \cite{SOEK94}.
These techniques are more involved from 
the computational point of view,
and we did not try to use them. They may, however, remedy
some of the difficulties found in our applications.

The simplest and the most straightforward way to calculate the
two-body effective interaction is accomplished by solving the equation
(\ref{eigomega}) and constructing the effective hamiltonian
according to Eq.(\ref{effomega}). The easiest way to perform
the calculations is in the center-of-mass and relative coordinate
basis and afterwards to do the transformation 
to the two-nucleon harmonic-oscillator basis. 
Note that $\omega$ is diagonal in the 
center-of-mass quantum numbers ${\cal N},{\cal L}$ 
as well as in $S,j,J,T$. 
Consequently, the sum in Eq.(\ref{eigomega})
goes only over $n_P, l_P$ for the basis states classified by
$|n l S j, {\cal N L}, J T\rangle$. An important point is the right
choice of the eigenstates in the set ${\cal K}$. For each 
$S,j,{\cal N},{\cal L},J,T$
we must choose as many relative coordinate eigenstates as the allowed
number of $n_P,l_P$ combinations. These are determined from
the condition $2{\cal N}+{\cal L}+2n_P+l_P\le N_{\rm max}$. 
Since the harmonic-oscillator basis is infinite, 
we make a truncation in the Q-space
by keeping only the states with $n_Q\le 150$. Note that $l=j\pm 1$
for the coupled channels and $l=j$ for the uncoupled channels.
This method can be easily applied to calculate the effective
interaction up to $N_{\rm max}=8$, as required in the present
shell-model calculations.

To take partially into account the many-body effects
neglected when using only a two-body effective interaction,
we employ the recently introduced multi-valued effective
interaction approach \cite{ZBVHS}. In this approach
different effective interactions are used 
for different $\hbar\Omega$ excitations.
The effective interactions then carry an additional index 
indicating the sum of the oscillator quanta for the spectators,
$N_{\rm sps}$, defined by
\begin{equation}\label{Nsps}
N_{\rm sps} = N_{\rm sum} - N_{\alpha}
= N'_{\rm sum} - N_{\gamma} \; ,
\end{equation}
where $N_{\rm sum}$ and $N'_{\rm sum}$ are the total oscillator
quanta in the initial and final many-body states, respectively, 
and $N_{\alpha}$
and $N_{\gamma}$ are the total oscillator quanta in the initial
and final two-nucleon states $|\alpha\rangle$ and $|\gamma\rangle$,
respectively.
Different sets of the effective interaction are determined
for different model spaces characterized by $N_{\rm sps}$ 
and defined by projection operators
\begin{mathletters}\label{projop}\begin{eqnarray}
Q_2(N_{\rm sps})&=&\left\{ 
\begin{array}{ll}
0 &  \mbox{if  $N_1+N_2\leq N_{\rm max} - N_{\rm sps}$} \; , \\ 
1 &  \mbox{otherwise} \; ;
\end{array}
\right.   \\
P_2(N_{\rm sps}) &=&
1-Q_2(N_{\rm sps}) \; .
\end{eqnarray}\end{mathletters}
This multi-valued effective-interaction approach is superior
to the traditional single-valued effective interaction,
as confirmed also in a model calculation \cite{NB96}.

The shell-model diagonalization is performed by using the
Many-Fermion-Dynamics Shell-Model Code \cite{VZ94}. We present
the results for $A=3-6$ nuclei in Tables \ref{tab1}-\ref{tab5}.
It has been observed earlier \cite{ZBVM} and it is apparent 
in the present calculations
as well that, when the effective interaction is derived using
the Lee-Suzuki method, the interaction becomes too strong and leads
to overbinding of nuclei. The problem is not in the calculational
method but rather in the fact that the original 
two-nucleon hamiltonian (\ref{hamomega2}) 
is flawed if $P_2\neq 1$. 
The difficulty is that while the relative-coordinate
harmonic-oscillator auxiliary potential is exactly cancelled
in the equation (\ref{hamomega}), it is not fully cancelled
when the effective interaction is derived from the two-nucleon
hamiltonian (\ref{hamomega2}). This can be seen, for example, 
when the 
nucleon-nucleon interaction is switched off. Then from 
Eq.(\ref{hamomega2}) we obtain an effective interaction
derived from the relative-coordinate harmonic-oscillator
potential, which will be different from the model-space part
of the harmonic-oscillator potential, appearing in 
Eq.(\ref{phamomegabeta}). Clearly, the $Q_2$-space part of the 
relative-coordinate harmonic-oscillator potential is responsible for
this effect. In order to reduce this spurious effect of the auxiliary
potential on the effective interaction, we introduce an {\it ad hoc}
modification of the relative-coordinate two-nucleon Q-space
part of the auxiliary potential as follows
\begin{eqnarray}\label{hamomega2rm}
H^\Omega_{2{\rm rel}}&=&\frac{\vec{q}^2}{2\mu}
+P_2\frac{A-2}{2A}\mu\Omega^2 \vec{r}^2P_2
+P_2k_Q\frac{A-2}{2A}\mu\Omega^2 \vec{r}^2Q_2
\nonumber \\
&&+Q_2k_Q\frac{A-2}{2A}\mu\Omega^2 \vec{r}^2P_2
+Q_2k_Q\frac{A-2}{2A}\mu\Omega^2 \vec{r}^2Q_2
\nonumber \\
&&+(1-k_Q)\frac{1}{2}\hbar\Omega\sqrt{\frac{A-2}{A}}(N_{\rm max}+2)Q_2
+ V(\vec{r}) \; .
\end{eqnarray}
Here we have introduced a constant $k_Q\le 1$; for $k_Q=1$ we get the
original hamiltonian appearing in Eq.(\ref{hamomega2r}). 
Moreover, for $A=2$ the harmonic-oscillator dependence vanishes
and for $P\rightarrow 1$ the relative-coordinate part of 
the hamiltonian (\ref{hamomega2r}) is recovered.  
In Eq.(\ref{hamomega2rm}) the Q-space harmonic-oscillator potential
is made more shallow and is shifted. The shift is such that
the P-space and the Q-space potentials are equal for 
$\vec{r}^2=\frac{1}{2} (\langle \vec{r}^2\rangle_{N_{\rm max}}
+\langle \vec{r}^2\rangle_{N_{\rm max}+1})$, with this mean value 
determined for the eigenstates of the relative-coordinate 
harmonic-oscillator hamiltonian appearing in Eq. (\ref{hamomega2r}). 
In this modified hamiltonian (\ref{hamomega2rm}),
a quasi-particle scattered into the Q-space feels a weaker auxiliary
potential. Note that an alternative method by
Krenciglowa {\it et al.} \cite{KKKO76} for deriving the G-matrix,
as opposed to that we employ here \cite{BHM71}, uses
plane-wave-type intermediate Q-space states, that is, no auxiliary
potential at all. For a recent review of this approach see 
Ref. \cite{HJKO95}. Our present modification, in fact, brings those two
methods closer together.
We solve the Schr\"{o}dinger equation with 
the hamiltonian (\ref{hamomega2rm}) by diagonalization in a
harmonic-oscillator basis characterized by 
$b=\sqrt{\frac{\hbar}{\mu\omega}}$ with the radial quantum number
$n=0\ldots 150$. The error caused by this truncation of the 
harmonic-oscillator basis can be estimated 
for $k_Q=1$, when the system can be solved as a differential equation. 
We found that the low-lying eigenvalues
obtained in the two calculations do not differ by more than 
$10^{-3}$ MeV and in most cases by much less. 
The lowest eigenvalues are typically of the order of 
$10^1$ MeV. The diagonalization cannot be used for hard-core 
nucleon-nucleon potentials, but for the soft-core Reid 93 
and Nijmegen II
potentials the interaction matrix elements can be evaluated
straightforwardly. Note that the hamiltonian (\ref{hamomega2rm})
depends on $N_{\rm max}$ and the projection operators. 
When the multi-valued effective interaction
is calculated, solutions are found only for projectors
and $N_{\rm max}$ corresponding to $N_{\rm sps}=0$. 
The value $N_{\rm max}=8$ is used in most calculations.
Using these solutions, all required sets of effective interactions
are constructed. 

The question arises if the modification we introduced in 
Eq. (\ref{hamomega2rm}) may not cause center-of-mass spurious 
contamination of the physical states.
We note that the modification is done 
only in the Q-space part of the auxiliary potential and,
most importantly, with regard to the two-nucleon relative 
coordinate. In conjuction with this, our choice
of the two-nucleon model space as well as the complete 
$N\hbar\Omega$ model space for the many-nucleon states 
prevents the center-of-mass contamination of physical states. 
We have tested numerically for possible spurious center-of-mass 
contaminations by varying the parameter $\beta$, 
introduced in Eq.(\ref{phamomegabeta}), in calculations 
for several systems and found that the physical states
remain unchanged for different choices of $\beta$, including
$\beta=0$, even if $k_Q\ne 1$.

We note that the free (or bare) values of the nucleon 
charges are used for calculating 
mean values of different operators.

In Table \ref{tab1} we present results of the 8$\hbar\Omega$
calculation for $^3$H with $\hbar\Omega=19.2$ MeV as suggested
by the formula \cite{B72} (in units of MeV):
\begin{equation}\label{hbaromega}
\hbar\Omega=45 A^{-\frac{1}{3}}-25 A^{-\frac{2}{3}} \; .
\end{equation}
We observe that the calculation, using the effective interaction
derived from the hamiltonian (\ref{hamomega2}), or equivalently
$k_Q=1$ in Eq. (\ref{hamomega2rm}), 
overbinds the system in comparison with both the experimental value
(8.482 MeV) and the exact result for the Reid 93 calculation 
(7.63 MeV) \cite{MSS96}.
Our aim should be to reproduce the result of the exact calculation. 
To achieve that
we varied the parameter $k_Q$ in Eq. (\ref{hamomega2rm}).
The value $k_Q=0.6$ gives reasonable agreement. We keep this
value for all other calculations, so that a meaningful comparison
of binding energies can be obtained.

In Table \ref{tab2} the calculated results for $^4$He are presented, 
where we have used 8$\hbar\Omega$ for the 
positive-parity states and 7$\hbar\Omega$ for the negative
parity states. 
The value $\hbar\Omega=18.4$ MeV, obtained from (\ref{hbaromega})
is employed.
The results for calculations with $k_Q=1$ and $k_Q=0.6$,
using the Reid 93 potential are shown, as well as 
calculations with $k_Q=0.6$, using the Nijmegen II potential.
We observe that both potentials give almost identical results.
The effective interaction derived using 
the unmodified relative-coordinate two-body hamiltonian
overbinds the system. On the other hand, the calculation
with the modified hamiltonian gives very reasonable agreement
with the experimental results. Unlike the experimental spectrum 
we still get 
the 2$\hbar\Omega$ $0^+$ state above the 1$\hbar\Omega$ $0^-$;
however, the discrepancy is reduced considerably in comparison
with previous calculations \cite{ZVB} and most of the states
compare better with the experiment data than in the recent 
calculation with multi-valued effective interaction \cite{ZBVHS}.

The calculation for $^5$He was performed for
several values of $\hbar\Omega$ and different
model spaces in order to study the
dependence of different states on these quantities. 
In Table \ref{tab3}
we show the 7$\hbar\Omega$ ($\pi=+$) 
and 6$\hbar\Omega$ ($\pi=-$) results, respectively, 
using $\hbar\Omega=17.8$ MeV obtained from 
(\ref{hbaromega}). The controversy regarding 
the shell-model calculations
for this nucleus has to do with the nature of the excited states 
\cite{CL96}. In the standard shell-model formulation, 
also employed here,
the center-of-mass of the nucleus is bound in a harmonic-oscillator
potential, so all states are bound. However, they do not necessarily
belong to the internal excitations of the studied nucleus. 
They may as well correspond to, e.g., a two-cluster configuration.
One would expect that such states are more sensitive to the
variation of the model space size and the harmonic-oscillator
binding potential \cite{ZVB96}. In Fig. \ref{fighe58} we present
the $\hbar\Omega$ dependence of the $^5$He states calculated
in the 7$\hbar\Omega$ ($\pi=+$) and 6$\hbar\Omega$ ($\pi=-$)
model spaces, respectively, by using 
the effective interaction obtained for $k_Q=0.6$ and $N_{\rm max}=8$
in Eq.(\ref{hamomega2rm}).
Note that only three experimental
states are known in this nucleus. Also the excitation energy
of the $\frac{1}{2}^-$ state is determined with a large error.
For comparison, in Fig. \ref{fighe56} results for the 5$\hbar\Omega$ 
($\pi=+$) and 4$\hbar\Omega$ ($\pi=-$) model spaces, respectively, 
are presented. The effective interaction for this calculation
has been derived from Eq.(\ref{hamomega2rm}) with the same $k_Q=0.6$
but different $N_{\rm max}=6$. We observe large sensitivity 
for the higher excited states on changes in $\hbar\Omega$ with the
exception of $\frac{3}{2}^+_2$. This state, dominated by the
$(0s)^3(0p)^2$ configuration, is a good candidate for the experimental
$\frac{3}{2}^+$ state. Note also the significant shifts 
of the excited states, again with the exception of $\frac{3}{2}^+_2$
and $\frac{1}{2}^-_1$, downwards in energy, when 
the model space is enlarged. Even if we cannot 
draw a definitive conclusion,
these facts indicate that excited states obtained
in the calculation, but unobserved, may be states,
which do not correspond to $^5$He internal excitations.

The calculated results for $^6$Li are presented in Table \ref{tab4}.
A 6$\hbar\Omega$ calculation was performed for the 
positive-parity states using $\hbar\Omega=17.2$ MeV obtained from
(\ref{hbaromega}). 
Again the effective interaction, derived
from the hamiltonian (\ref{hamomega2}), is too strong and
overbinds the system. The calculation with the modified
hamiltonian (\ref{hamomega2rm}) gives very reasonable agreement
with experiment for all properties shown.  In addition to the 
experimental states
presented in the Table \ref{tab4}, another $J^\pi=3^+,T=0$ state
with the excitation energy 15.8 Mev is known. 
The lowest $J^\pi=3^+,T=0$
2$\hbar\Omega$ state obtained in our $k_Q=0.6$ calculation
has an energy of 13.840 MeV, with a main configuration 
of $(0s)^4(0p)^1(1p0f)^1$. 
Another candidate for this experimental
state may be the calculated state with an excitation energy 
of 18.378 MeV, dominated by the configurations $(0s)^4(1s0d)^2$ and
$(0s)^3(0p)^2(1s0d)^1$. Likely, the latter state may
be more stable against model space and $\hbar\Omega$
variations. We also performed a 5$\hbar\Omega$ calculation
for the negative-parity states of $^6$Li. The lowest
calculated state is $J^\pi=2^-,T=0$ with an excitation energy
9.135 MeV, followed by $J^\pi=1^-,T=0$ ($E_x$=9.335 MeV) and
$J^\pi=0^-,T=0$ ($E_x$=11.372 MeV). Such states have not been
observed experimentaly.

In Table \ref{tab5} we present the $^6$He characteristics.
As isospin symmetry is not broken  in the present
calculations, the energies are the same as those 
for $^6$Li with $T=1$. In addition, the proton and neutron radii
are evaluated. Reasonable agreement with the experimantal
values is found. 

Contrary to the previous
no-core shell-model calculations \cite{ZVB,ZBVHS}, the effective
interaction used in the present calculations does not contain
a parameter $\Delta$
adjusted for each nucleus in order to get a reasonable binding
energy. Besides the model-space size,
the present calculations depend only 
on the harmonic-oscillator frequency $\hbar\Omega$
and on the parameter $k_Q$, appearing in the relative coordinate
two-nucleon hamiltonian (\ref{hamomega2rm}), which distinguishes
the structure of the P-space from that of the Q-space. For the results
presented in Tables \ref{tab1}-\ref{tab5}, we chose $\hbar\Omega$
according to formula (\ref{hbaromega}) and $k_Q$ was fixed
to reproduce the result of the exact $^3$H calculation. Consequently, 
our calculations contain no variable parameters and meaningful 
comparisons can be made between our calculated
binding energies and the quantities derived from them 
and the experimental values as well as the results of 
other calculations. 

In this regard, we note the recent calculation of valence energies of
$^6$He and $^6$Li and the neutron separation energy of $^5$He, 
using the two-frequency shell-model approach \cite{KMN96}. 
For example, our
calculation with $k_Q=0.6$ and the Reid 93 potential gives
for the neutron separation energy of $^5$He,
$E_{\rm sp}=E_B(^5{\rm He})-E_B(^4{\rm He})=-1.303$ MeV 
to be compared with the experimental value of
$-0.894$ MeV. Similarly, the valence energy of $^6$He is found to be 
\( E_{\rm val}(^6{\rm He})=-[E_B(^6{\rm He})+E_B(^4{\rm He})
-2 E_B(^5{\rm He})]=-3.151\; {\rm MeV}\), 
compared with the experimental $-2.761$ MeV. Furthermore,
\( E_{\rm val}(^6{\rm Li})
=-[E_B(^6{\rm Li})+E_B(^4{\rm He})
-E_B(^5{\rm He})-E_B(^5{\rm Li})]=-6.845\; {\rm MeV}\), 
where we used $E_B(^5{\rm Li})=E_B(^5{\rm He})=26.105$ MeV 
from Table \ref{tab3}, as the Coulomb contributions, not taken into
account in the calculations, more or less cancel. 
Here the experimental value is $-6.559$ MeV. We observe good agreement
of these quantities with the experimental values, in fact, better than
that obtained in Ref. \cite{KMN96}. Apparently, 
the reason for these improved results is the size of the
model space, which is much larger in our calculations. 

In Ref. \cite{CL96} concerns were raised about the stability
of $^6$Li to the $\alpha+d$ threshold in previous no-core 
calculations \cite{ZVB}. From the present results we deduce
that $^6$Li is bound against this threshold by 1.314 MeV,
compared with the experimental value of 1.474 MeV. Note that
the deuteron is treated exactly in our formalism, 
$E_B(d)=2.2246$ MeV. To arrive at the present number,
we used the Coulomb energy contributions to binding energy
$E_C(Z\equiv 2)=-0.76$ MeV and $E_C(Z\equiv 3)=-1.46$ MeV.
Another issue raised in Ref. \cite{CL96} was sign of 
the quadrupole moment of $^6$Li.
As in the previous no-core calculations
\cite{ZVB,ZBVHS}, we also get the sign correctly in 
our $k_Q=0.6$ calculation.
Clearly, this is a consequence  of a reasonable effective interaction
and a large enough model space.

\section{Conclusions}
\label{sec4}

In the present paper we have presented different ways of
constructing a starting-energy-independent two-body 
effective interaction. In particular, we studied the Lee-Suzuki
similarity transformation method and compared
two different iteration schemes for obtaining a solution, 
the vertex renormalization method
and the Krenciglowa-Kuo method. When applying the
vertex renormalization method, we obtained the derivatives
of the G-matrix exactly by using the derivatives of the 
reference G-matrix. A third approach involved a direct calculation
of the transformation operator $\omega$. When the convergence
conditions of the iteration methods are satisfied, all
three approaches lead to the same two-body effective
interaction. Our conclusion is that, for large
model spaces that we employ,  the only viable option
is the direct calculation method. For large model
spaces ($N_{\rm max}>4$), the iteration methods 
either fail to converge or the calculations are
time consuming or both. 

Unlike in most previous applications,
where the non-hermitian effective interaction was hermitized by
averaging the conjugate matrix elements, we hermitize
the effective interaction exactly using a similarity transformation. 
It was demonstrated in Ref. \cite{KEHZSOK} that the averaging is a good
approximation for the hermitian $sd$ and $pf$ effective interactions. 
We observe here, however, that for large model spaces, like those
we employ, the non-hermiticity could be significant. It is certainly
preferable to work with the exactly hermitized interaction.

Employing the derived effective interaction, we have 
performed shell-model calculations, in which 
up to 9 major harmonic-oscillator
shells may be occupied. In order to take into account
part of the many-body effects, we utilized the new multi-valued
effective interaction approach \cite{ZBVHS}. 
The results for no-core, full $N\hbar\Omega$
calculations are given for nuclei with $A=3-6$.
The effective interaction was derived from the Reid 93
nucleon-nucleon potential, but calculations were also made 
with Nijmegen II potential for comparison. 
As observed earlier \cite{ZBVM},
when the Lee-Suzuki method is applied for a two-nucleon
system with a harmonic-oscillator auxiliary potential,
the resulting effective interaction is too strong
and leads to overbinding of the many-body system.
This problem is caused by incomplete cancellation of the relative
coordinate part of the auxiliary potential.
To mend this flaw, we introduced a modification
of the Q-space part of the relative-coordinate
two-nucleon hamiltonian, from which the effective
interaction is calculated. In effect this weakens the
auxiliary potential in the Q-space. 

In the present calculations, besides the model space size,
the only free parameters are the harmonic-oscillator
frequency $\hbar\Omega$ and the parameter modifing the
two-nucleon hamiltonian, as discussed above. The latter
parameter was fixed from the $^3$H binding energy calculation
(fitted to the result of an exact $^3$H calculation)
and $\hbar\Omega$ was taken from the phenomenological
formula (\ref{hbaromega}). Hence, unlike   
previous no-core calculations, we are able to compare,
for nuclei other than $^3$H, quantities
derived from binding energies, such as valence energies, 
with experiment. 

For most calculated characteristics
we found good agreement with the experimental values.
In agreement with the previous observation \cite{ZB94}, 
we found that the Reid 93 and Nijmegen II potentials
give very similar results. The question of 
low-lying positive-parity states in $^5$He was also
investigated. We observed that the calculated low-lying 
positive-parity states have
not converged, regarding changes in the model-space 
size and variations in $\hbar\Omega$. This may indicate that
they do not correspond to internal excitations
of $^5$He. However, we are not in a position to make
a conclusive statement concerning this issue.

Because we derive the transformation
operator $\omega$ explicitly in our calculations, 
we are able to use it
for calculating any effective operator, employing the approaches
discussed in Refs. \cite{NGK93,OS}, in a similar way as
in our model calculations \cite{NB96}.
Besides calculating other effective operators, 
we also intend to extend the calculations to larger $A$.

\acknowledgements{
This work was supported by the NSF grant No. PHY93-21668.
P.N. also acknowledges partial support from 
the Czech Republic grant GA ASCR A1048504.
}

\begin{figure}
\caption{Energy dependence for the $^5$He states 
on the harmonic-oscillator energy $\hbar\Omega$
in a full 6 and 7 $\hbar \Omega$ calculation.
Only the three lowest negative-parity states are shown.
The positive-parity states $\frac{3}{2}^+, \frac{5}{2}^+$ and
$\frac{7}{2}^+, \frac{9}{2}^+$ have very close energies. Therefore,
only single lines for these pairs of states are presented.
The second $\frac{3}{2}^+$ state is dominated by the $s^3 p^2$ 
configuration and should correspond to the experimental  
$\frac{3}{2}^+$ state. For further description of the calculation 
see the text.
}
\label{fighe58}
\end{figure}

\begin{figure}
\caption{Energy dependence for the $^5$He states 
on the harmonic-oscillator energy $\hbar\Omega$
in a full 4 and 5 $\hbar \Omega$ calculation.
See Fig.~\protect{\ref{fighe58}} for the details.
}
\label{fighe56}
\end{figure}

\begin{table}
\begin{tabular}{cccc}
$^3$H & $k_Q=1$ & $k_Q=0.6$ & Exp    \\
\hline
$E_B$ &8.739 & 7.674&8.482  \\
$\sqrt{\langle r_p^2\rangle}$ &1.540 &1.600 &1.41-1.62 \\
$\mu$ &2.638 &2.634 &2.979 
\end{tabular}
\caption{Experimental and calculated binding energy in MeV, 
point proton radius in fm, and
magnetic moment in $\mu_{\rm N}$, for $^3$H.   
The ground state has $J^\pi=\frac{1}{2}^+$. 
The 8$\hbar\Omega$ calculation
results are presented. Different $k_Q$ choices correspond
to different two-nucleon hamiltonians, from which the
multi-valued effective interaction is calculated, as
explained in the text. The Reid 93 nucleon-nucleon potential 
is used, and the harmonic-oscillator
parameter is taken to be $\hbar\Omega=19.2$ MeV. 
The exact, calculated binding energy for this potential 
is 7.63 MeV. The free nucleon $g$ factors were used. 
The experimental values are taken from Ref. 
\protect\cite{TWH87}.}
\label{tab1}
\end{table}

\begin{table}
\begin{tabular}{ccccc}
$^4$He & Reid 93&Reid 93 & Nijmegen II& Exp\\
 & $k_Q=1$ & $k_Q=0.6$ &$k_Q=0.6$&    \\
\hline
$E_B$          & 31.115 & 27.408 & 27.499 & 28.296 \\
$\sqrt{\langle r_p^2\rangle}$ &1.378 &1.434 &1.432 &1.46 \\
$E_x(0^+_1,0)$ & 0      & 0      & 0      & 0     \\
$E_x(0^+_2,0)$ & 24.009 & 21.619 & 21.724 & 20.21 \\
$E_x(0^-_1,0)$ & 23.506 & 21.290 & 21.402 & 21.01 \\
$E_x(2^-_1,0)$ & 25.118 & 22.852 & 22.948 & 21.84 \\
$E_x(2^-_1,1)$ & 26.544 & 24.056 & 24.125 & 23.33 \\
$E_x(1^-_1,1)$ & 26.874 & 24.263 & 24.325 & 23.64 \\
$E_x(1^-_1,0)$ & 27.763 & 25.113 & 25.205 & 24.25 \\
$E_x(0^-_1,1)$ & 27.940 & 25.247 & 25.325 & 25.28 \\
$E_x(1^-_2,1)$ & 28.154 & 25.419 & 25.487 & 25.95 
\end{tabular}
\caption{Experimental and calculated binding energy, 
point proton radius in fm,
and excitation energies $E_x(J^\pi,T)$ for $^4$He.   
All energies are in MeV.
The 8$\hbar\Omega$ $(\pi=+)$ and 7$\hbar\Omega$ $(\pi=-)$ calculation
results are presented. Different $k_Q$ choices correspond
to different two-nucleon hamiltonians, from which the
multi-valued effective interaction was calculated, as
explained in the text. Both the Reid 93 and Nijmegen II
nucleon-nucleon potentials are used for comparison purposes. 
The harmonic-oscillator parameter is taken to be $\hbar\Omega=18.4$ MeV.
The experimental values are taken from Refs. 
\protect\cite{TWH92,VJV87}.}
\label{tab2}
\end{table}

\begin{table}
\begin{tabular}{cccc}
$^5$He & $k_Q=1$ & $k_Q=0.6$ & Exp    \\
\hline
$E_B$          & 30.454 & 26.105 & 27.402 \\
$\sqrt{\langle r_p^2\rangle}$ &1.531 &1.597 & N/A \\
$\mu$ &-1.845  &-1.844  & N/A \\ 
$Q$ &-0.437  &-0.489  & N/A   \\
$E_x(\frac{3}{2}^-_1,\frac{1}{2})$ & 0     & 0     & 0   \\
$E_x(\frac{1}{2}^-_1,\frac{1}{2})$ & 2.537 & 2.026 & 4$\pm$1  \\
$E_x(\frac{1}{2}^+_1,\frac{1}{2})$ & 4.062 & 3.113 & N/A  \\
$E_x(\frac{3}{2}^+_1,\frac{1}{2})$ & 9.503 & 8.099 & N/A  \\
$E_x(\frac{5}{2}^+_1,\frac{1}{2})$ & 9.520 & 8.172 & N/A  \\ 
$E_x(\frac{3}{2}^-_2,\frac{1}{2})$ & 11.200 & 10.413 & N/A  \\
$E_x(\frac{1}{2}^-_2,\frac{1}{2})$ & 15.049 & 14.025 & N/A  \\
$E_x(\frac{1}{2}^+_2,\frac{1}{2})$ & 20.931 & 18.652 & N/A  \\ 
$E_x(\frac{3}{2}^+_2,\frac{1}{2})$ & 21.378 & 19.454 & 16.75   
\end{tabular}
\caption{Experimental and calculated binding energy, 
point proton radius in fm,
magnetic moment in $\mu_{\rm N}$,
quadrupole moment in $e$fm$^2$
and excitation energies $E_x(J^\pi,T)$ for $^5$He.   
All energies are in MeV.
The 7$\hbar\Omega$ $(\pi=+)$ and 6$\hbar\Omega$ $(\pi=-)$ 
calculation
results are presented. Different $k_Q$ choices correspond
to different two-nucleon hamiltonians, from which the
multi-valued effective interaction was calculated, as
explained in the text. The Reid 93 
nucleon-nucleon potential is used, and the harmonic-oscillator
parameter is taken to be $\hbar\Omega=17.8$ MeV.
The calculated $\frac{3}{2}^+_2$ state is associated
with the experimental $\frac{3}{2}^+$ state as it is dominated
by $(0s)^3(0p)^2$ configuration. See also figures 
\protect{\ref{fighe58}} and \protect{\ref{fighe56}}.
The experimental values are taken from Ref. 
\protect\cite{AS88}.}
\label{tab3}
\end{table}

\begin{table}
\begin{tabular}{cccc}
$^6$Li & $k_Q=1$ & $k_Q=0.6$ & Exp    \\
\hline
$E_B$          & 37.532 & 31.647 & 31.995 \\
$\sqrt{\langle r_p^2\rangle}$ &1.995 &2.097 & 2.42 \\
$\mu$ & 0.839  & 0.839  & 0.822  \\ 
$Q$ & 0.027  & -0.052  & -0.082  \\
$E_x(1^+_1,0)$ & 0     & 0     & 0   \\
$E_x(3^+_1,0)$ & 2.368 & 2.398 & 2.19   \\
$E_x(0^+_1,1)$ & 4.301 & 3.695 & 3.56   \\
$E_x(2^+_1,0)$ & 4.966 & 4.438 & 4.31   \\
$E_x(2^+_1,1)$ & 6.878 & 6.144 & 5.37   \\
$E_x(1^+_2,0)$ & 7.232 & 6.277 & 5.65   \\
$E_x(2^+_2,1)$ & 10.641& 9.333 & N/A   \\
$E_x(1^+_1,1)$ & 11.442& 10.019& N/A   \\
$E_x(1^+_3,0)$ & 11.847& 10.467& N/A   \\
$E_x(0^+_2,1)$ & 13.866& 12.050& N/A   
\end{tabular}
\caption{Experimental and calculated binding energy, 
point proton radius in fm,
magnetic moment in $\mu_{\rm N}$,
quadrupole moment in $e$fm$^2$,
and excitation energies $E_x(J^\pi,T)$ for $^6$Li.   
All energies are in MeV.
The 6$\hbar\Omega$ calculation
results are presented. Different $k_Q$ choices correspond
to different two-nucleon hamiltonians, from which the
multi-valued effective interaction was calculated, as
explained in the text. The Reid 93 
nucleon-nucleon potential is used, and the harmonic-oscillator
parameter is chosen to be $\hbar\Omega=17.2$ MeV.
Only the states dominated by the $0\hbar\Omega$ configuration
are presented.
See the text for the discussion of other states. 
The experimental values are taken from Refs. 
\protect\cite{VJV87,AS88}.}
\label{tab4}
\end{table}

\begin{table}
\begin{tabular}{cccc}
$^6$He & $k_Q=1$ & $k_Q=0.6$ & Exp    \\
\hline
$E_B$          & 33.230 & 27.953 & 29.269 \\
$\sqrt{\langle r_p^2\rangle}$ &1.629 & 1.707 & 1.72 \\
$\sqrt{\langle r_n^2\rangle}$ &2.196 & 2.317 & 2.59  \\
$E_x(0^+_1,1)$ & 0     & 0     & 0   \\
$E_x(2^+_1,1)$ & 2.577 & 2.449 & 1.80  \\
$E_x(2^+_2,1)$ & 6.339 & 5.638 & N/A   \\
$E_x(1^+_1,1)$ & 7.140 & 6.324 & N/A   \\
$E_x(0^+_2,1)$ & 9.565 & 8.355 & N/A   
\end{tabular}
\caption{Experimental and calculated binding energy, 
point proton and neutron radius in fm,
and excitation energies $E_x(J^\pi,T)$ for $^6$He.   
All energies are in MeV.
The presented results are for 6$\hbar\Omega$ and are performed
as described in Table \protect\ref{tab4}.
The experimental values are taken from Refs. 
\protect\cite{AS88,T93}.}
\label{tab5}
\end{table}


\begin{references}

\bibitem {ZBJVC} D.C. Zheng, B.R. Barrett, L. Jaqua, J.P. Vary, and
      R.J. McCarthy, Phys. Rev. {\bf C 48}, 1083 (1993).

\bibitem {JZBV} L. Jaqua, D.C. Zheng, B.R. Barrett, and J.P. Vary,
                Phy. Rev. {\bf C 48}, 1765 (1993).
	 
\bibitem {JHBV} L. Jaqua, P. Halse, B.R. Barrett, and J.P. Vary,
                Nucl. Phys. A{\bf 571}, 242 (1994).

\bibitem {ZBVC} D.C. Zheng, B.R. Barrett, J.P. Vary, and
      R.J. McCarthy, Phys. Rev. {\bf C 49}, 1999 (1994).

\bibitem {ZB94} D.C. Zheng and B.R. Barrett, Phys. Rev. 
         {\bf C 49}, 3342 (1994).

\bibitem {ZVB} D.C. Zheng, J.P. Vary, and B.R. Barrett, Phys. Rev. 
               {\bf C 50}, 2841 (1994).
         
\bibitem {ZBVM} D.C. Zheng, B.R. Barrett, J.P. Vary, and
      H. M\"{u}ther, Phys. Rev. {\bf C 51}, 2471 (1995).
	 
\bibitem {ZBVHS} D.C. Zheng, B.R. Barrett, J.P. Vary, 
                 W.C. Haxton, and C.L. Song, Phys. Rev. 
                 {\bf C 52}, 2488 (1995).

\bibitem {NB96} P.Navr\'atil, and B.R. Barrett, 
             Phys. Lett. {\bf B 369}, 193 (1996).

\bibitem {BHM71} B.R. Barrett, R.G.L. Hewitt, and R.J. McCarthy,
             Phys. Rev. {\bf C 3}, 1137 (1971).

\bibitem {LS80} K. Suzuki and S.Y. Lee, Prog. Theor. Phys. {\bf 64}, 
                2091 (1980).

\bibitem {KK} E.M. Krenciglowa and T.T.S. Kuo, Nucl. Phys. 
              {\bf A 235}, 171 (1974).

\bibitem {KKSO} T.T.S. Kuo, F. Krmpoti\'{c}, K. Suzuki, R. Okamoto,
             Nucl. Phys. {\bf A 582}, 205 (1995).

\bibitem {PB91} N.A.F.M. Popppelier and P.J. Brussaard, Nucl. Phys.
            {\bf A 530}, 1 (1991).

\bibitem {MWC93} M.R. Meder, S.W. Walker, and B.R. Caldwell,
              Nucl. Phys. {\bf A556}, 228 (1993).

\bibitem {S82SO83} K. Suzuki, Prog. Theor. Phys. {\bf 68},
              246 (1982); 
            K. Suzuki and R. Okamoto, Prog. Theor. Phys. {\bf 70},
              439 (1983).

\bibitem {SGH} F.G. Scholtz, H.B. Geyer, and F.J.W. Hahne,
             Ann. Phys. (NY) {\bf 213}, 74 (1992).

\bibitem {SKTS} V.G.J. Stoks, R.A.M. Klomp, C.P.F. Terheggen, 
             and J.J. de Swart, Phys. Rev. {\bf C 49} 2950 (1994).

\bibitem {Spc} V.G.J. Stoks, private communication.

\bibitem {NG93} P.Navr\'atil and H.B. Geyer, Nucl. Phys. {\bf A 556},
                165 (1993).

\bibitem {SOEK94} K. Suzuki, R. Okamoto, P.J. Ellis, T.T.S. Kuo,
            Nucl. Phys. {\bf A567}, 576 (1994). 

\bibitem {VZ94} J.P. Vary and D.C. Zheng, ``The Many-Fermion-Dynamics
            Shell-Model Code'', Iowa State University (1994)
            (unpublished).

\bibitem {TWH87} D.T. Tilley, H.R. Weller, and H.H. Hasan, Nucl. Phys.
            {\bf A474}, 1 (1987).

\bibitem {TWH92} D.T. Tilley, H.R. Weller, and G.M. Hale, Nucl. Phys.
             {\bf A541}, 1 (1992).

\bibitem {VJV87} H. De Vries, C.W. De Jager, and C. De Vries,
            At. Data and Nucl. Data Tables {\bf 36}, 495 (1987).

\bibitem {AS88} F. Ajzenberg-Selove, Nucl. Phys. {\bf A490},
              1 (1988).

\bibitem {T93} I. Tanihata, D. Hirata, T. Kobayashi, S. Shimoura,
               K. Sugimoto, and H. Toki, Phys. Lett. {\bf B 289},
               261 (1992).

\bibitem {KKKO76} E.M. Krenciglowa, C.L. Kung, T.T.S. Kuo,
             and E. Osnes, Ann. Phys. (NY) {\bf 101}, 154 (1976).

\bibitem {HJKO95} M. Hjorth-Jensen, T.T.S. Kuo, E. Osnes,
          Phys. Rep. {\bf 261}, 125 (1995).

\bibitem {B72} G.F. Bertsch, ``The Practioner's Shell Model'',
               North Holland (1972).

\bibitem {MSS96} R. Machleidt, F. Sammarruca, and Y. Song,
            Phys. Rv. C {\bf 53}, R1483 (1996).

\bibitem {CL96} Attila Cs\'ot\'o and Rezs\"{o} G. Lovas, Phys. Rev. C
            {\bf 53}, 1444 (1996).

\bibitem {ZVB96} D.C. Zheng, J.P. Vary, and B.R. Barrett, 
           Phys. Rev. C {\bf 53}, 1447 (1996).

\bibitem {KMN96} T.T.S. Kuo, H. M\"{u}ther, and K. Amir-Amir-Nili,
         ``Realistic effective interactions for halo nuclei'',
         preprint, (1996).

\bibitem {KEHZSOK} T.T.S. Kuo, P.J. Ellis, Jifa Hao, Zibang Li,
             K. Suzuki, R. Okamoto,
             and H. Kumagai, Nucl. Phys. {\bf A560}, 621 (1993).

\bibitem {NGK93} P.Navr\'atil, H.B. Geyer and T.T.S. Kuo, 
             Phys. Lett. {\bf B 315}, 1 (1993).

\bibitem {OS} K. Suzuki and R. Okamoto, Prog. Theor. Phys. {\bf 93},
              905 (1995).



\end{references}
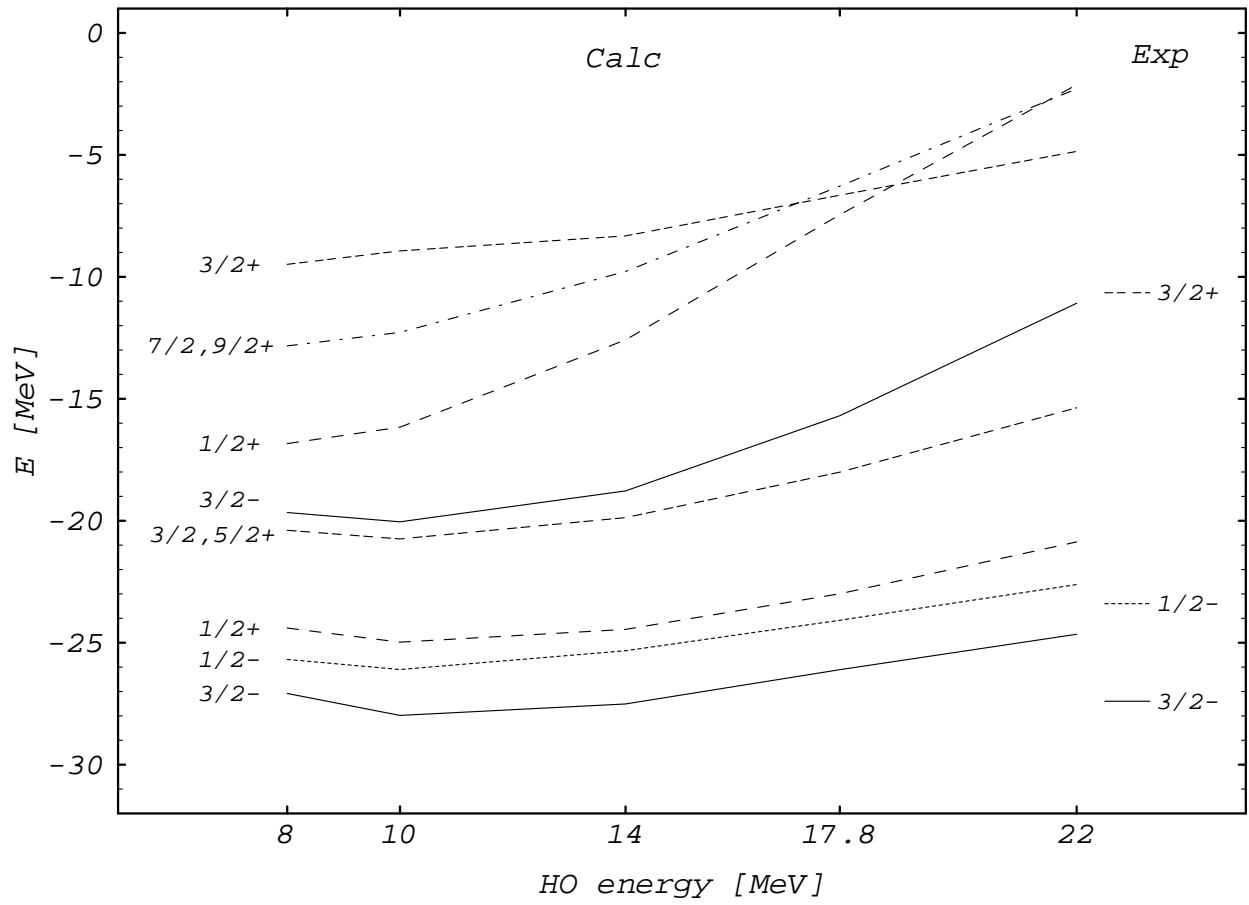
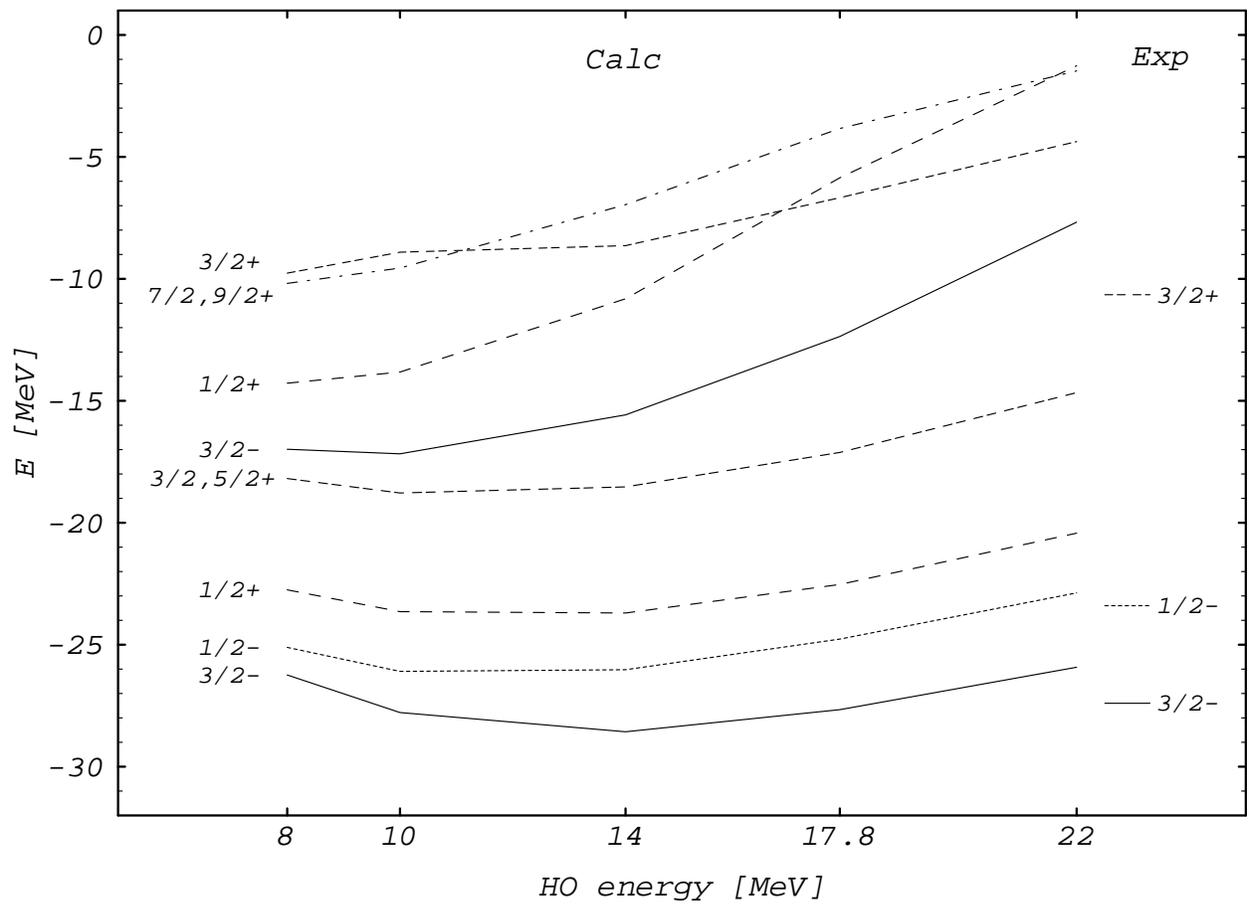
\end{document}